\documentclass[%
superscriptaddress,
preprint,
nofootinbib,
nobibnotes,
 amsmath,amssymb,
 aps,
]{revtex4-1}

\usepackage{graphicx}
\usepackage{dcolumn}
\usepackage{amsmath}
\usepackage{textcomp}

\newcommand{\vect}[1]{\boldsymbol{\mathbf{#1}}}

\begin{document}

\title{Weak anti-localization of two-dimensional holes in germanium beyond the diffusive regime}

\author{C.-T. Chou}
\affiliation{Department of Electrical Engineering, National Taiwan University, Taipei 10617, Taiwan}
\author{N. T. Jacobson}
\affiliation{Sandia National Laboratories, Albuquerque, New Mexico 87185, USA}
\author{J. E. Moussa}
\affiliation{Sandia National Laboratories, Albuquerque, New Mexico 87185, USA}
\author{A. D. Baczewski}
\affiliation{Sandia National Laboratories, Albuquerque, New Mexico 87185, USA}
\author{Y. Chuang}
\affiliation{Graduate Institute of Electronic Engineering, National Taiwan University, Taipei 10617, Taiwan}
\author{C.-Y. Liu}
\affiliation{Graduate Institute of Electronic Engineering, National Taiwan University, Taipei 10617, Taiwan}
\author{J.-Y. Li}
\email{jiunyun@ntu.edu.tw}
\affiliation{Graduate Institute of Electronic Engineering, National Taiwan University, Taipei 10617, Taiwan}
\affiliation{National Nano Device Laboratories, Hsinchu 30078, Taiwan}
\affiliation{Department of Electrical Engineering, National Taiwan University, Taipei 10617, Taiwan}
\author{T. M. Lu}
\email{tlu@sandia.gov}
\affiliation{Sandia National Laboratories, Albuquerque, New Mexico 87185, USA}

\date{\today}

\begin{abstract}
Gate-controllable spin-orbit coupling is often one requisite for spintronic devices.  
For practical spin field-effect transistors, another essential requirement is ballistic spin transport, where the spin precession length is shorter than the mean free path such that the gate-controlled spin precession is not randomized by disorder.  
In this letter, we report the observation of a gate-induced crossover from weak localization to weak anti-localization in the magneto-resistance of a high-mobility two-dimensional hole gas in a strained germanium quantum well.  
From the magneto-resistance, we extract the phase-coherence time, spin-orbit precession time, spin-orbit energy splitting, and cubic Rashba coefficient over a wide density range.  
The mobility and the mean free path increase with increasing hole density, while the spin precession length decreases due to increasingly stronger spin-orbit coupling.
As the density becomes larger than $\sim6\times 10^{11}$cm$^{-2}$, the spin precession length becomes shorter than the mean free path, and the system enters the ballistic spin transport regime.
We also report here the numerical methods and code developed for calculating the magneto-resistance in the ballistic regime, where the commonly used HLN and ILP models for analyzing weak localization and anti-localization are not valid.
These results pave the way toward silicon-compatible spintronic devices.  

\end{abstract}

\maketitle

Spin-orbit coupling (SOC) in low-dimensional semiconductor systems has received much attention for its importance in both fundamental studies and spintronic applications.  
For example, it is the underlying physical mechanism giving rise to the spin Hall effect \cite{Hirsch99} and the quantum spin Hall effect \cite{Kane95}, and is also one essential ingredient for creating Majorana zero modes in conventional semiconductors \cite{Sau2010, Lutchyn10,Alicea12}.  
For spintronic devices, SOC not only provides a means to control the rotation of carrier spins \cite{Datta90} but also lifts the degeneracy of the two spin states, enabling all-electric spin-selecting nanostructures \cite{Koga02,Debray09}.

SOC in a two-dimensional (2D) system can be seen as a $\vect{k_{||}}$-dependent effective magnetic field induced by inversion asymmetry, where $\vect{k_{||}}$ is the in-plane wave vector.  
Carrier spins precess about the effective magnetic field axis with a precession frequency $|\vect{\Omega_3}|$ determined by the strength of SOC.  
There are two types of SOC in a semiconductor heterostructure, the Rashba SOC and the Dresselhaus SOC.  
The former is caused by structural inversion asymmetry, typically along the growth direction of the thin-film heterostructure.  
In addition to the built-in structural asymmetry, the Rashba SOC can also be controlled by external electric fields through electrostatic gating.  
The Dresselhaus SOC exists in crystals with bulk inversion asymmetry, such as those with the Zincblende structure.  
In spintronic devices, such as the spin field-effect transistor (FET) proposed by Datta and Das \cite{Datta90}, the Rashba SOC is the more relevant mechanism, owing to the gate tunability \cite{Nitta97}.  
We also note that in Ge, the material system of interest in this work, the Dresselhaus SOC can be ignored because of the bulk inversion symmetry in this material.

In a 2D hole gas (2DHG), the Rashba SOC is cubic in $\vect{k_{||}}$ in the spin-orbit Hamiltonian ($\vect{H_{SO}}$) due to the nature of the heavy hole band \cite{Winkler02,Bi13,Moriya14}:
\begin{equation} \label{eq:Hamiltonian}
  \vect{H_{SO}} = \hbar\vect{\sigma}\cdot\vect{\Omega_3} = \alpha_3E_zi\left(k_-^3\sigma_+ - k_+^3\sigma_- \right)
\end{equation}

Here $\hbar$ is the reduced Planck constant, $\vect{\sigma}$ is the Pauli vector, $\vect{\Omega_3}$ is the precession frequency due to the cubic Rashba SOC, $\alpha_3$ is the cubic Rashba coefficient, $E_z$ is the effective electric field along the $z$ direction, $k_\pm=k_x\pm ik_y$ where $k_x$ and $k_y$ are the components of $\vect{k_{||}}$, and $\sigma_\pm=(\sigma_x\pm\sigma_y)/2$ where $\sigma_x$ and $\sigma_y$ are the Pauli matrices. The details of cubic Rashba SOC are reviewed in Ref. \onlinecite{Winkler08}.

Measurement of anomalous magneto-resistance due to weak localization (WL) and weak anti-localization (WAL) is one common method to study the SOC in low-dimensional systems.  
The WL effect is a positive resistance correction at low magnetic fields due to the quantum interference of time-reversed paths taken by the carriers.  
The WAL effect, on the other hand, is a negative resistance correction at low magnetic fields caused by SOC in the system.  
SOC introduces a Berry phase of $\pi$ between two time reversed paths and reverses the sign of the resistance correction \cite{BERGMANN84}.  
The introduction of magnetic field breaks the time reversal symmetry and washes out the WL and WAL effect.  
The strength of SOC can be quantitatively characterized by extracting the spin precession length and time, together with phase coherence length and time, from the magneto-resistance using magneto-transport models developed for the WL and WAL effects.  
The most commonly used models are the HLN model proposed by Hikami {\it et al.} \cite{hikami1980} and the ILP model proposed by Iordanskii  {\it et al.} \cite{iordanskii1994}.

Despite most studies of SOC focused on III-V materials \cite{Dresselhaus55,Dresselhaus92,Heida98,Grundler00,Koga02b,Miller03,Wunderlich05,Studer09,vanWeperen15}, recent studies have demonstrated considerable SOC strength in Ge 2DHGs \cite{Moriya14, Morrison14, Failla15, Foronda15,Mizokuchi17}.  
There are two advantages of using Ge 2DHGs for spintronic applications.  
First, the Dresselhaus SOC is absent in a Ge 2DHG system, leaving the system purely governed by the tunable Rashba SOC.  
The second, and perhaps more important, advantage of Ge is its compatibility with modern complementary-metal-oxide-semiconductor (CMOS) technology.  
The controllability and scalability enabled by the unmatched CMOS technology make Ge a promising material candidate for spintronic devices.

\begin{figure}[h]
\resizebox{6.69 in}{!}{\includegraphics{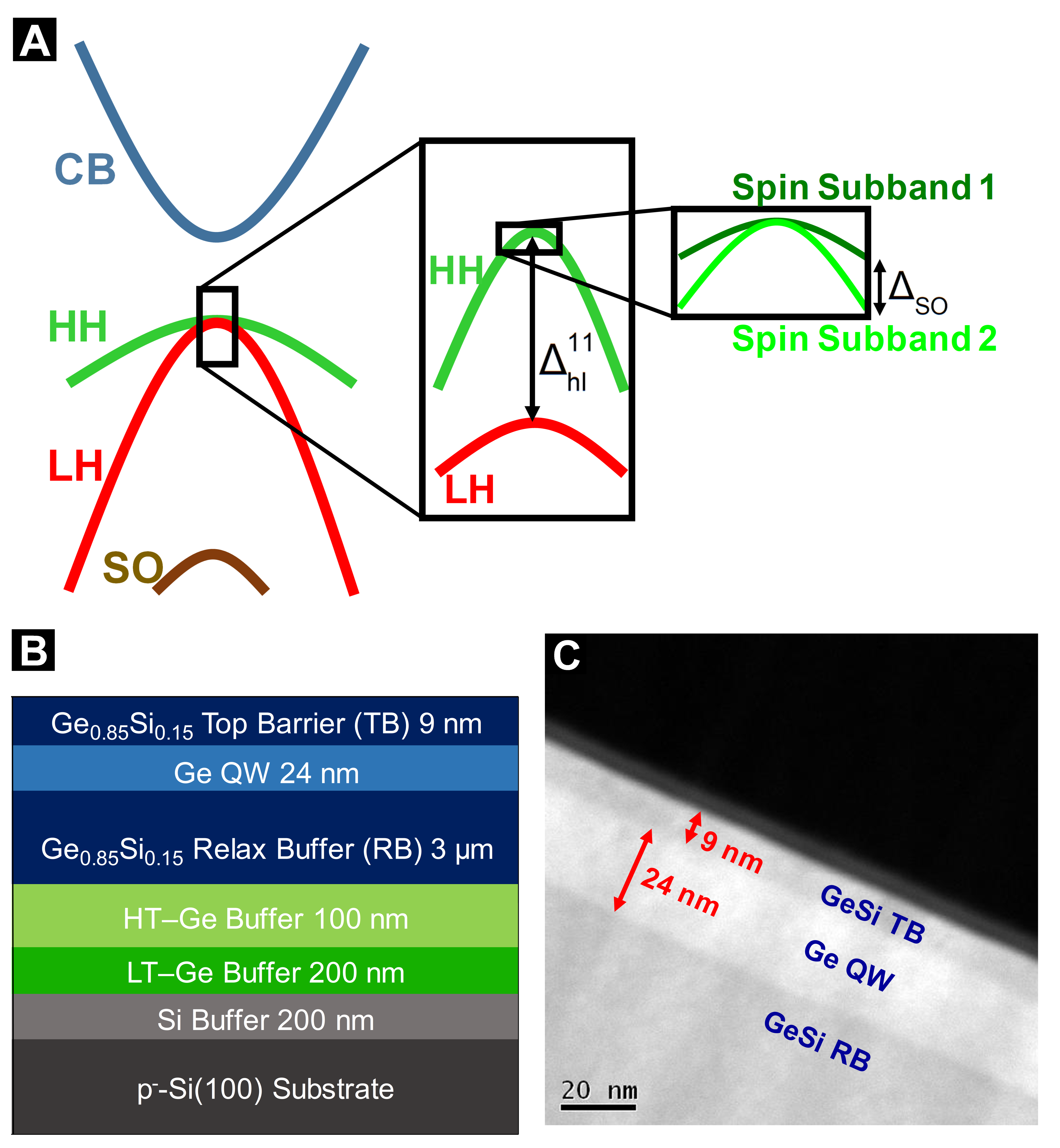}}
\caption{\label{Fig1} (\textbf{A}) The valence band structure  of a compressively strained Ge QW.  The degenerate LH and HH bands in the bulk band structure, shown in the left, split with an energy spacing of $\Delta_{hl}^{11}$ because of strain and quantum confinement, as shown in the middle.  SOC further splits the HH band into two spin subband at finite $\vect{k_{||}}$, as shown in the right.  (\textbf{B}) Schematic cross-section and (\textbf{C}) XTEM image of the epitaxial heterostructure. }
\end{figure}

A schematic of the valence band structure of Ge is shown in Fig.~1\textbf{A}.  
In bulk Ge, the six-fold degenerate valence band is lifted by SOC, forming a two-fold degenerate spin-orbit band (SO), a two-fold degenerate light hole band (LH), and a two-fold degenerate heavy hole band (HH), as shown in the left.  
The LH and HH bands are degenerate at $\vect{k_{||}}=0$.  
In a (100)-oriented compressively strained Ge quantum well (QW), this LH-HH degeneracy is lifted by strain and quantum confinement \cite{Moriya14}, as shown in the middle.  
The HH band is the lower-energy band for holes, while the LH band is higher in energy by $\Delta_{hl}^{11}$.
The HH band is further split by the cubic Rashba SOC in Ge into two spin sub-bands with an energy splitting $\Delta_{SO}$, as shown in the right.

Most of the previous studies on the SOC in Ge 2DHGs were carried out using modulation-doped heterostructures \cite{Moriya14, Morrison14, Failla15, Foronda15}. 
However, those structures suffer from either parallel conduction in the doping layer \cite{Morrison14, Foronda15} or very low mobility \cite{Moriya14}.  
Furthermore, the doping layer above the 2DHG could screen out the electric field from the top gate, resulting in limited gate tunability.  
This limited gate tunability is a significant constraint for spintronic applications, where gating is used not only to control the Rashba SOC but also to implement spin filtering \cite{Chuang2015}.  
Using the undoped heterostructure FET (HFET) architecture allows us to circumvent these problems.  
We have previously reported the realization of high-mobility 2DHGs with very wide density ranges in Ge HFETs \cite{Su2017}.  
In particular, the density in a shallow Ge HFET with a QW-to-surface distance of  9 nm can be set as high as $7.5\times 10^{11}$ cm$^{-2}$ with a mobility of $6.1\times 10^{4}$ cm$^{2}$V$^{-1}$s$^{-1}$.  
In this work, we use this undoped Ge/GeSi heterostructure to study the SOC of high-mobility 2DHGs in a Ge QW through magneto-resistance measurements.  
We observe a crossover from WL to WAL as the hole density increases.  
The phase-coherence time, spin-orbit precession time, spin-orbit energy splitting, and cubic Rashba coefficient are extracted. 
The density dependence of these parameters shows that the Rashba SOC is widely tunable and that the 2DHG system enters the ballistic spin transport regime at high densities, two essential requirements for realizing Ge-based spin FETs.

An undoped Ge/GeSi heterostructure was grown on a Si (100) wafer by reduced pressure chemical vapor deposition with GeH$_4$ and SiH$_4$ as the precursors.  
First, 200 nm of Si followed by 200 nm relaxed Ge were grown on top of a Si (100) wafer.  
High-temperature in-situ annealing was then performed at 825$^{\circ}$C.  
On top of the Ge layer, the following layers were epitaxially grown in order: 100 nm of Ge, 3 $\mu$m of Ge$_{0.85}$Si$_{0.15}$, 24 nm of Ge, and 9 nm of Ge$_{0.85}$Si$_{0.15}$.  
The epitaxial layer structure is shown in Fig.~1\textbf{B}.  
The details of the growth and hole transport behavior were reported in our prior work \cite{Su2017}.  
The thicknesses of the Ge QW and GeSi top barrier layers were confirmed to be 24 and 9 nm, respectively, by cross-sectional transmission electron microscopy (XTEM), as shown in Fig.~1\textbf{C}.

Ge HFETs were fabricated using standard photolithography.  
Al was deposited by electron beam evaporation followed by lift-off.  
Rapid thermal annealing was performed to create Ohmic contacts.  
Then, 60 nm of Al$_2$O$_3$ was deposited using atomic layer deposition, followed by Ti/Au deposition for the gate layer.  
Electrical access to the Ohmic contacts was made by etching away Al$_2$O$_3$ and deposition of metal pads.
Low-field  magneto-resistance was measured at 260 mK in a $^3$He cryogenic system using standard low-frequency lock-in techniques with the hole density modulated by varying the gate bias.

\begin{figure}[h]
\resizebox{6.69 in}{!}{\includegraphics{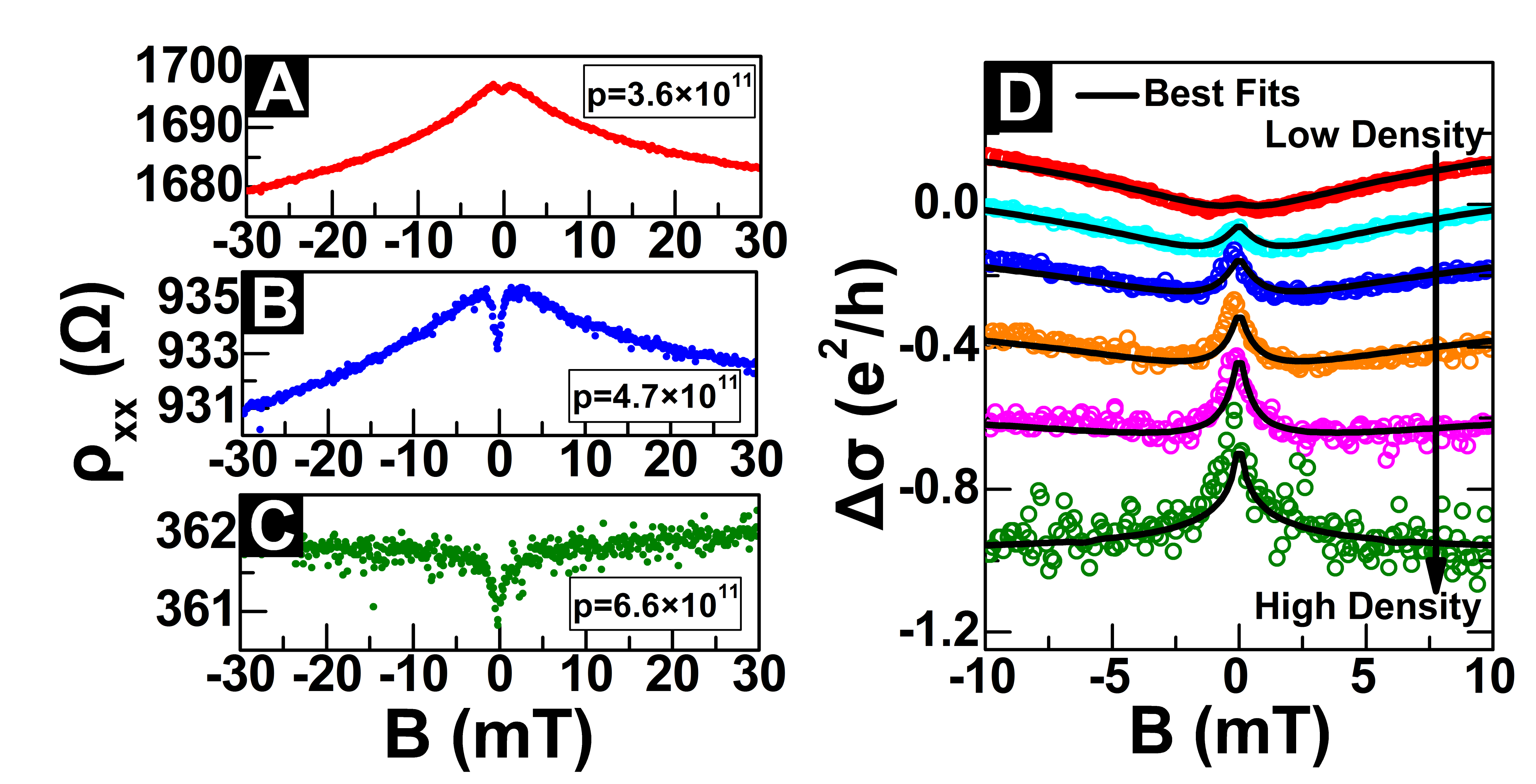}}
\caption{\label{Fig2} The magneto-resistance at densities $p=$ (\textbf{A}) 3.6, (\textbf{B}) 4.7, and (\textbf{C}) 6.6 $\times$10$^{11}$ cm$^{-2}$, respectively.   (\textbf{D}) The magneto-conductance $\Delta\sigma$ at $p=$ 3.6, 4.1, 4.7, 5.1, 5.5, and 6.6$\times$10$^{11}$ cm$^{-2}$, from top to bottom.  The black lines represent the best fit curves from G\&G's model \cite{Glazov2006}.  The curves are shifted vertically for clarity.}
\end{figure}

Figures 2\textbf{A}, 2\textbf{B}, and 2\textbf{C} show the longitudinal magneto-resistance $\rho_{xx}(B)$ at densities of 3.6, 4.7, and 6.6 $\times$10$^{11}$ cm$^{-2}$, respectively.
In the low-density regime (Fig.~2\textbf{A}), the resistance decreases mostly monotonically with the magnetic field, which is known as the WL effect.  
A weak resistance dip at zero magnetic field due to the WAL effect is barely visible.  
As the hole density is increased by the gate voltage, the zero-field resistance dip becomes stronger, with a persisting broad background from the WL effect (Fig.~2\textbf{B}).  
At an even higher density, only the WAL peak remains and no WL effect is observed (Fig.~2\textbf{C}).

Since the WAL effect is caused by the SOC in the 2DHG system, the observed crossover from the WL-dominant regime to the WAL-dominant regime is evidence that the SOC is gate tunable.  
The WAL peak becomes more pronounced as the hole density increases, showing that the SOC effect is stronger at higher hole densities. 
This is consistent with the Rashba SOC.  
The increased bias voltage provides larger structural asymmetry through a larger electric field $E_z$.
Furthermore, the higher density corresponds to a larger Fermi wave vector $k_F$. 
Both the larger $E_z$ and $k_F$ lead to stronger SOC.

To be more quantitative, one can extract the spin coherence length and the spin precession frequency from the magneto-resistance.  
The most common method of extracting the SOC parameters is to fit the magneto-resistance curves to the HLN model \cite{hikami1980,Dresselhaus92} or the ILP model \cite{iordanskii1994,Nakamura12,Liang15}. 
However, for those two models to be applicable, the following two criteria need to be satisfied: (i) the system is in the diffusive spin transport regime, where the spin precession length $L_{SO}$ is much larger than the mean free path $L_{tr}$, and (ii) the applied magnetic field is smaller than the transport characteristic magnetic field $B_{tr}$, defined as $\hbar/2eL_{tr}^2$.  
In our high-mobility 2DHG system, $L_{tr}$ ranges from 0.2 $\mu$m to 1 $\mu$m, and $B_{tr}$ ranges from 0.1 to 10 mT.  
The measured magneto-resistance in this work is mostly outside the parameter space where the HLN and the ILP models are valid.  
To extract the spin-orbit parameters from our data, we use a more general formalism developed by Glazov and Golub (G\&G) \cite{Glazov2006}, which is valid beyond the two criteria and is applicable to our data.  
The nontrivial part in performing this analysis is to properly evaluate the numerical integrals for magneto-resistance corrections.  
In the Supplementary Information, we include a discussion on how to properly perform the numerical integral calculation as well as the details of the fitting process.  
Also included in the Supplementary Information is the code for performing such calculations.
In Fig.~2\textbf{D} we show the magneto-conductivity $\Delta\sigma(B)$, defined as $\sigma(B)-\sigma(0)$, for a series of densities and the best fit curves.  
We can see that G\&G's model fits the measured magneto-resistance reasonably well.

\begin{figure}[h]
\resizebox{6.69 in}{!}{\includegraphics{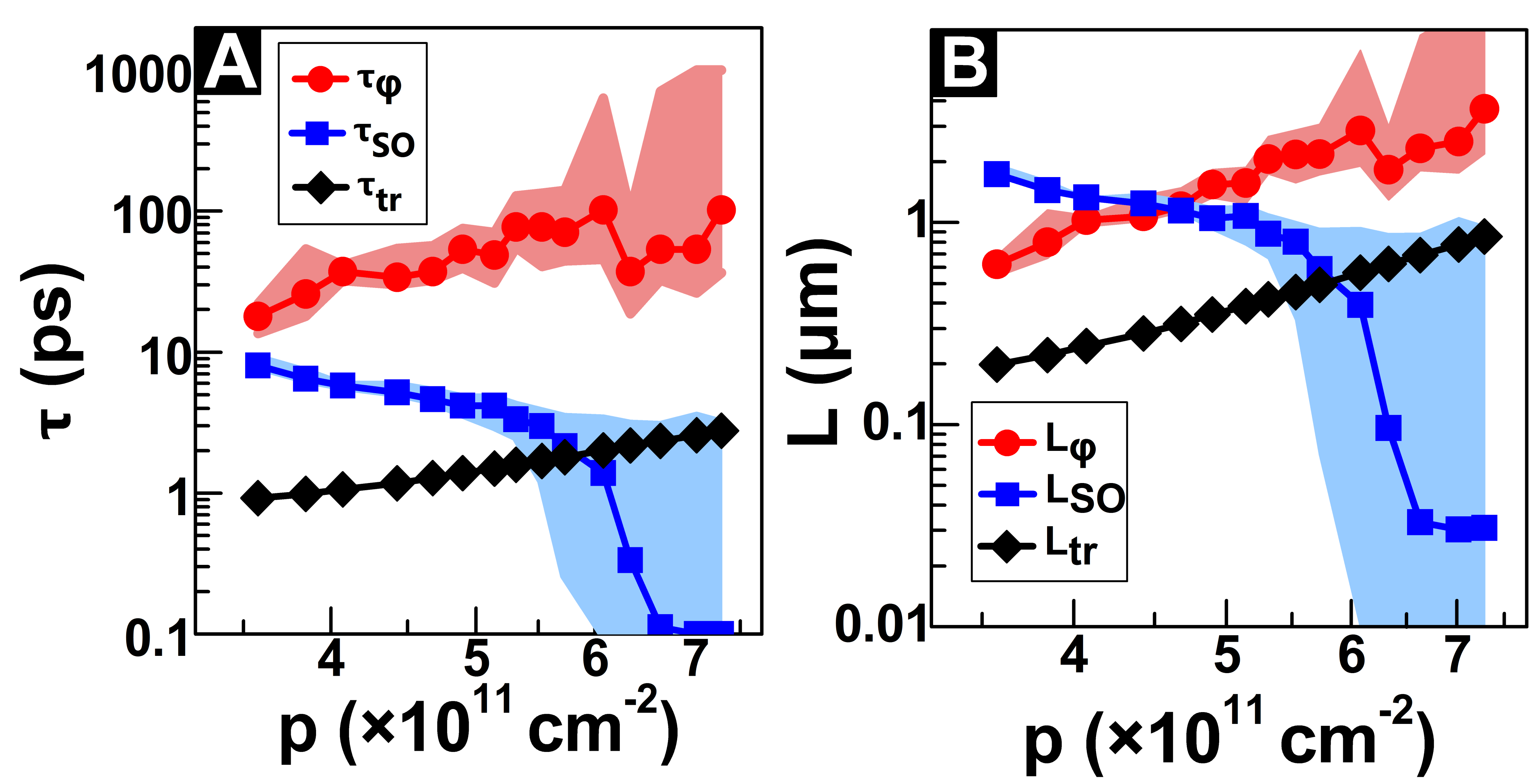}}
\caption{\label{Fig3} (\textbf{A}) Time and (\textbf{B}) the corresponding length scales obtained from the curve fitting of magneto-resistance.  The subscripts $SO$, $\phi$, and $tr$ indicate spin-orbit, phase coherence, and transport.  The uncertainty of fitting parameters is indicated by the shaded regions.}
\end{figure}

At each density, we extract two parameters from the fitting: the spin-orbit precession frequency $\Omega_3$ and the phase coherence time $\tau_\phi$.  
The spin precession time $\tau_{SO}$ is defined as $1/\Omega_3$, and the transport scattering time $\tau_{tr}$ is defined as $(m^* \mu)/e$, where $m^*=0.08m_0$ is the hole effective mass in the 2D plane, obtained from temperature-dependent Shubnikov-de Haas (SdH) oscillations.  
$\tau_{SO}$, $\tau_\phi$, and $\tau_{tr}$ are shown in Fig.~3\textbf{A}.  
Figure 3\textbf{B} shows the three corresponding length scales, the spin precession length $L_{SO}=\tau_{SO}\times v_F$, the mean free path $L_{tr}=\tau_{tr}\times v_F$, and the phase coherence length $L_\phi=\sqrt{D\tau_\phi}$, where $v_F$ is the Fermi velocity and $D$ is the diffusion constant.  
As shown in Fig.~3(\textbf{B}), the spin precession length decreases monotonically with density.  
This is consistent with the argument earlier that the SOC in our 2DHG system become stronger as the density increases. 
Due to the nature of the WAL effect, in the regime where the WL background is absent, the magneto-resistance curve is mostly determined by $\tau_\phi$, and fitting only provides an upper bound of $\tau_{SO}$ \cite{Grbic08}.

The results in Fig.~\ref{Fig3} demonstrate the potential of using Ge 2DHG for spintronic applications.  
First, at $p>6\times 10^{11}$cm$^{-2}$, the 2DHG enters the ballistic spin regime, where $\tau_{SO}<\tau_{tr}<\tau_\phi$.  
This allows holes to transport ballistically through a channel while the spin precesses \cite{sugahara2010}.  
Furthermore, the tunability of the strength of Rashba SOC is a crucial property for building spintronic devices, for it allows for a direct control of the spin precession rate using electrostatic gating \cite{Datta90}. 
The tunability of Rashba SOC in the Ge 2DHG is demonstrated by the gate-induced change in $L_{SO}$, from $\sim$2 $\mu$m at $3.6\times 10^{11}$cm$^{-2}$ to below 0.1 $\mu$m at $6.4\times 10^{11}$cm$^{-2}$, as shown in Fig.~3\textbf{B}.  
The observation of tunable Rashba SOC in the ballistic spin regime paves the way toward CMOS-compatible spin FETs, where gating controls whether a spin is parallel or antiparallel with the magnetization of the spin detector \cite{koo2009}.

\begin{figure}[h]
\resizebox{6.69 in}{!}{\includegraphics{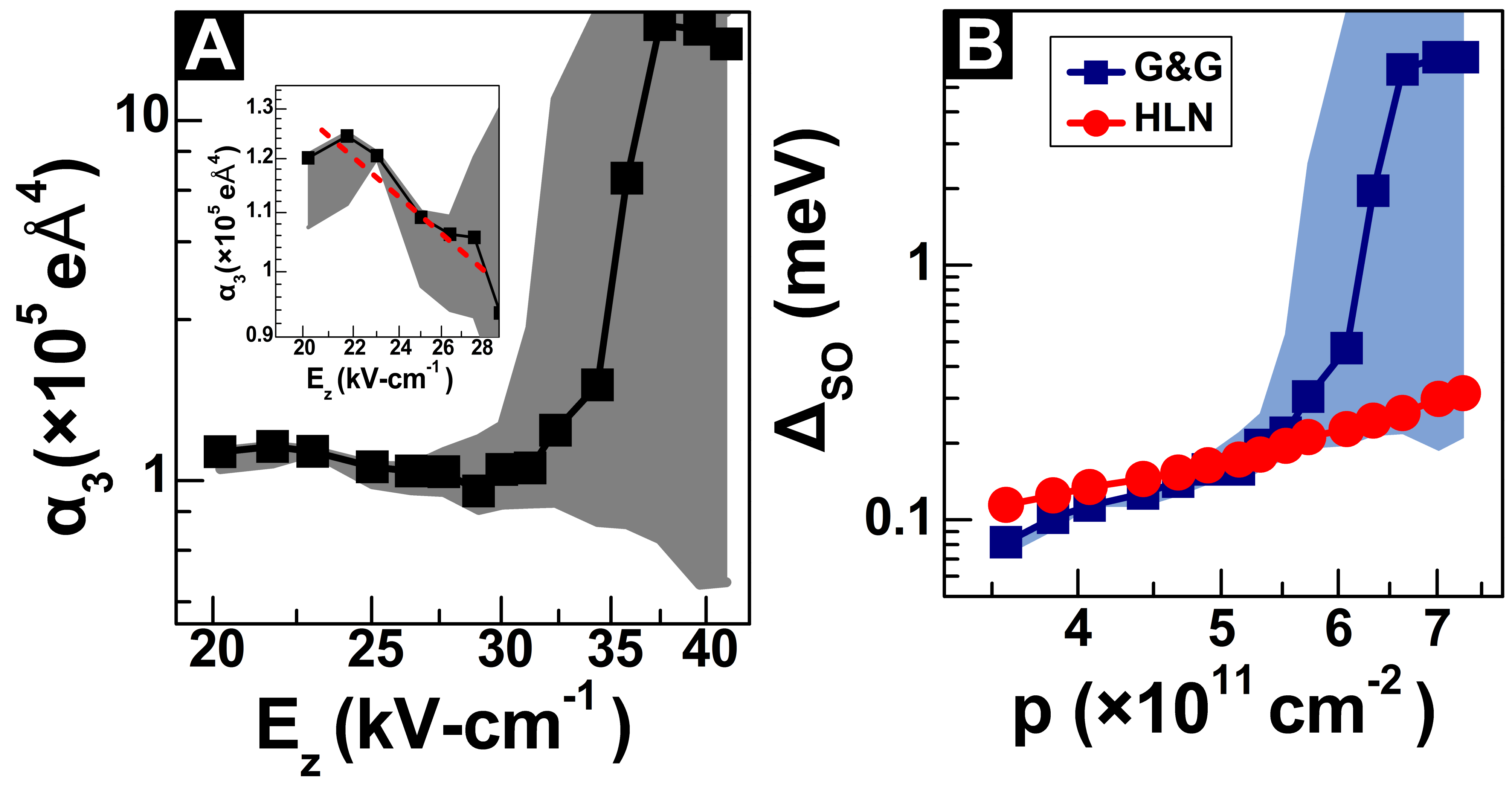}}
\caption{\label{Fig4} (\textbf{A}) The cubic Rashba coefficient ($\alpha_3$) as a function of the vertical electric field in a log-log plot.  The inset shows a zoom-in view at smaller electric fields.  The red dashed line indicates a power-law dependence with an exponent of -0.5.  (\textbf{B}) The energy splitting between the two spin sub-bands $\Delta_{SO}$ extracted from the parameters in Fig.~\ref{Fig3}. The uncertainty of fitting parameters is indicated by the shaded regions.}
\end{figure}

The cubic Rashba coefficient $\alpha_3$ and the spin-orbit energy splitting $\Delta_{SO}$ at the Fermi energy are shown in Fig.~\ref{Fig4}\textbf{A} and \textbf{B}, respectively. 
$\alpha_3$ and $\Delta_{SO}$ are calculated using the relation $\Delta_{SO}=\hbar|\Omega_3|=\alpha_3 〈E_z 〉 k_F^3$, where $〈E_z 〉$ is the average z-direction electric field \cite{Winkler2003book}.  
In the low-density regime where the error bars are small, the red dotted line shows that $\alpha_3$ decreases as $E_z$ increases.  
This counter-intuitive phenomenon has been reported experimentally \cite{Winkler02,Habib04} and explained theoretically \cite{Winkler2003book} by the change in LH-HH spin splitting energy with respect to $E_z$.  
A power-law dependence $\alpha_3\sim〈E_z 〉^{-4/3}$ is predicted in systems where the LH-HH splitting is dominated by quantum confinement in a triangular QW.  
On the other hand, $\alpha_3$ is expect to be invariant with respect to $〈E_z 〉$ when the LH-HH splitting is dominated by strain-induced energy splitting.  
A weak power-law with an exponent of $\sim$-0.5 observed in this work suggests that both quantum confinement and strain contribute to the LH-HH splitting.

The $\Delta_{SO}$ shown in in Fig.~\ref{Fig4}\textbf{B} is simply the rescaled spin-orbit precession frequency $\Omega_3$ for comparison purposes.  
It allows us to compare our results to the fitting results obtained by using the HLN model, shown as the red dots in Fig.~\ref{Fig4}.  
Although the criteria required by the HLN model are not satisfied, surprisingly the HLN model appears to fit the magneto-resistance reasonably well.  
Using the HLN model outside the regime where the model is valid, as is sometimes done in the literature \cite{Grbic08}, provides a reasonable estimate of the SOC strength.
In Table I, we compare our extracted Rashba coefficient with those reported by other groups for a Ge 2DHG system.  
Here we use $\beta_3=\alpha_3 〈E_z〉$ for comparison purposes.
While the mobility of the 2DHG system in different studies varies significantly, the values of $\beta_3$ are all comparable in magnitude, on the order of $1\times$10$^{-28}$ eVm$^3$.

\begin{table}

\begin{tabular}{| p{5cm} | p{2.5cm} | p{2.5cm} | p{2.5cm} | p{2.5cm} |}
\hline
Reference & This work & Ref. \onlinecite{Mizokuchi17} & Ref. \onlinecite{Moriya14} & Ref. \onlinecite{Foronda15}  \\  \hline
Structure & Undoped QW  & MOS & Modulation-doped QW& Modulation-doped QW\\ \hline
Extraction Method & WL-WAL & WL-WAL & WL-WAL & SdH \\ \hline
$\beta_3$ ($\times$10$^{-28}$ eVm$^3$) & 0.3 & 0.7 & 0.2 & 1.0 \\ \hline
Peak Mobility (cm$^2$V$^{-1}$s$^{-1}$) & 61,000	 & 4,000	 & 5,000	 & 780,000 \\ \hline

\end{tabular}
\caption{Comparison of the values of $\beta_3$ obtained from different groups for a Ge 2DHG system.  }
\end{table}

In conclusion, we observed a clear crossover from WL to WAL in magneto-resistance in an undoped Ge/GeSi 2DHG system.  
The gate tunability of Rashba SOC strength is demonstrated.  
Further analyses suggest that the system enters the ballistic spin transport regime at $p>6\times 10^{11}$cm$^{-2}$.  
The ballistic spin transport, combined with the tunable Rashba SOC, makes undoped Ge HFETs a promising architecture for CMOS-compatible spintronic devices.

This work has been supported by the Division of Materials Sciences and Engineering, Office of Basic Energy Sciences, U.S. Department of Energy (DOE).  This work was performed, in part, at the Center for Integrated Nanotechnologies, a U.S. DOE, Office of Basic Energy Sciences, user facility.  Sandia National Laboratories is a multi-mission laboratory managed and operated by National Technology and Engineering Solutions of Sandia, LLC., a wholly owned subsidiary of Honeywell International, Inc., for the U.S. Department of Energy's National Nuclear Security Administration under contract DE-NA-0003525. The views expressed in the article do not necessarily represent the views of the U.S. Department of Energy or the United States Government.  The work at National Taiwan University has been supported by the Ministry of Science and Technology under the contract 106-2112-M-002-009- and the Ministry of Education through the Higher Education Sprout Projects (career development project: NTU-107L7818 and NTU core consortium: 107L891702).

%\bibliography{Ge2DHGSOC}

\begin{thebibliography}{39}%
\makeatletter
\providecommand \@ifxundefined [1]{%
 \@ifx{#1\undefined}
}%
\providecommand \@ifnum [1]{%
 \ifnum #1\expandafter \@firstoftwo
 \else \expandafter \@secondoftwo
 \fi
}%
\providecommand \@ifx [1]{%
 \ifx #1\expandafter \@firstoftwo
 \else \expandafter \@secondoftwo
 \fi
}%
\providecommand \natexlab [1]{#1}%
\providecommand \enquote  [1]{``#1''}%
\providecommand \bibnamefont  [1]{#1}%
\providecommand \bibfnamefont [1]{#1}%
\providecommand \citenamefont [1]{#1}%
\providecommand \href@noop [0]{\@secondoftwo}%
\providecommand \href [0]{\begingroup \@sanitize@url \@href}%
\providecommand \@href[1]{\@@startlink{#1}\@@href}%
\providecommand \@@href[1]{\endgroup#1\@@endlink}%
\providecommand \@sanitize@url [0]{\catcode `\\12\catcode `\$12\catcode
  `\&12\catcode `\#12\catcode `\^12\catcode `\_12\catcode `\%12\relax}%
\providecommand \@@startlink[1]{}%
\providecommand \@@endlink[0]{}%
\providecommand \url  [0]{\begingroup\@sanitize@url \@url }%
\providecommand \@url [1]{\endgroup\@href {#1}{\urlprefix }}%
\providecommand \urlprefix  [0]{URL }%
\providecommand \Eprint [0]{\href }%
\providecommand \doibase [0]{http://dx.doi.org/}%
\providecommand \selectlanguage [0]{\@gobble}%
\providecommand \bibinfo  [0]{\@secondoftwo}%
\providecommand \bibfield  [0]{\@secondoftwo}%
\providecommand \translation [1]{[#1]}%
\providecommand \BibitemOpen [0]{}%
\providecommand \bibitemStop [0]{}%
\providecommand \bibitemNoStop [0]{.\EOS\space}%
\providecommand \EOS [0]{\spacefactor3000\relax}%
\providecommand \BibitemShut  [1]{\csname bibitem#1\endcsname}%
\let\auto@bib@innerbib\@empty
%</preamble>
\bibitem [{\citenamefont {Hirsch}(1999)}]{Hirsch99}%
  \BibitemOpen
  \bibfield  {author} {\bibinfo {author} {\bibfnamefont {J.~E.}\ \bibnamefont
  {Hirsch}},\ }\href {\doibase 10.1103/PhysRevLett.83.1834} {\bibfield
  {journal} {\bibinfo  {journal} {Phys. Rev. Lett.}\ }\textbf {\bibinfo
  {volume} {83}},\ \bibinfo {pages} {1834} (\bibinfo {year}
  {1999})}\BibitemShut {NoStop}%
\bibitem [{\citenamefont {Kane}\ and\ \citenamefont {Mele}(2005)}]{Kane95}%
  \BibitemOpen
  \bibfield  {author} {\bibinfo {author} {\bibfnamefont {C.~L.}\ \bibnamefont
  {Kane}}\ and\ \bibinfo {author} {\bibfnamefont {E.~J.}\ \bibnamefont
  {Mele}},\ }\href {\doibase 10.1103/PhysRevLett.95.226801} {\bibfield
  {journal} {\bibinfo  {journal} {Phys. Rev. Lett.}\ }\textbf {\bibinfo
  {volume} {95}},\ \bibinfo {pages} {226801} (\bibinfo {year}
  {2005})}\BibitemShut {NoStop}%
\bibitem [{\citenamefont {Sau}\ \emph {et~al.}(2010)\citenamefont {Sau},
  \citenamefont {Lutchyn}, \citenamefont {Tewari},\ and\ \citenamefont
  {Das~Sarma}}]{Sau2010}%
  \BibitemOpen
  \bibfield  {author} {\bibinfo {author} {\bibfnamefont {J.~D.}\ \bibnamefont
  {Sau}}, \bibinfo {author} {\bibfnamefont {R.~M.}\ \bibnamefont {Lutchyn}},
  \bibinfo {author} {\bibfnamefont {S.}~\bibnamefont {Tewari}}, \ and\ \bibinfo
  {author} {\bibfnamefont {S.}~\bibnamefont {Das~Sarma}},\ }\href {\doibase
  10.1103/PhysRevLett.104.040502} {\bibfield  {journal} {\bibinfo  {journal}
  {Phys. Rev. Lett.}\ }\textbf {\bibinfo {volume} {104}},\ \bibinfo {pages}
  {040502} (\bibinfo {year} {2010})}\BibitemShut {NoStop}%
\bibitem [{\citenamefont {Lutchyn}\ \emph {et~al.}(2010)\citenamefont
  {Lutchyn}, \citenamefont {Sau},\ and\ \citenamefont {Das~Sarma}}]{Lutchyn10}%
  \BibitemOpen
  \bibfield  {author} {\bibinfo {author} {\bibfnamefont {R.~M.}\ \bibnamefont
  {Lutchyn}}, \bibinfo {author} {\bibfnamefont {J.~D.}\ \bibnamefont {Sau}}, \
  and\ \bibinfo {author} {\bibfnamefont {S.}~\bibnamefont {Das~Sarma}},\ }\href
  {\doibase 10.1103/PhysRevLett.105.077001} {\bibfield  {journal} {\bibinfo
  {journal} {Phys. Rev. Lett.}\ }\textbf {\bibinfo {volume} {105}},\ \bibinfo
  {pages} {077001} (\bibinfo {year} {2010})}\BibitemShut {NoStop}%
\bibitem [{\citenamefont {Alicea}(2010)}]{Alicea12}%
  \BibitemOpen
  \bibfield  {author} {\bibinfo {author} {\bibfnamefont {J.}~\bibnamefont
  {Alicea}},\ }\href {\doibase 10.1103/PhysRevB.81.125318} {\bibfield
  {journal} {\bibinfo  {journal} {Phys. Rev. B}\ }\textbf {\bibinfo {volume}
  {81}},\ \bibinfo {pages} {125318} (\bibinfo {year} {2010})}\BibitemShut
  {NoStop}%
\bibitem [{\citenamefont {Datta}\ and\ \citenamefont {Das}(1990)}]{Datta90}%
  \BibitemOpen
  \bibfield  {author} {\bibinfo {author} {\bibfnamefont {S.}~\bibnamefont
  {Datta}}\ and\ \bibinfo {author} {\bibfnamefont {B.}~\bibnamefont {Das}},\
  }\href@noop {} {\bibfield  {journal} {\bibinfo  {journal} {Appl. Phys.
  Lett.}\ }\textbf {\bibinfo {volume} {56}},\ \bibinfo {pages} {665} (\bibinfo
  {year} {1990})}\BibitemShut {NoStop}%
\bibitem [{\citenamefont {Koga}\ \emph
  {et~al.}(2002{\natexlab{a}})\citenamefont {Koga}, \citenamefont {Nitta},
  \citenamefont {Takayanagi},\ and\ \citenamefont {Datta}}]{Koga02}%
  \BibitemOpen
  \bibfield  {author} {\bibinfo {author} {\bibfnamefont {T.}~\bibnamefont
  {Koga}}, \bibinfo {author} {\bibfnamefont {J.}~\bibnamefont {Nitta}},
  \bibinfo {author} {\bibfnamefont {H.}~\bibnamefont {Takayanagi}}, \ and\
  \bibinfo {author} {\bibfnamefont {S.}~\bibnamefont {Datta}},\ }\href
  {\doibase 10.1103/PhysRevLett.88.126601} {\bibfield  {journal} {\bibinfo
  {journal} {Phys. Rev. Lett.}\ }\textbf {\bibinfo {volume} {88}},\ \bibinfo
  {pages} {126601} (\bibinfo {year} {2002}{\natexlab{a}})}\BibitemShut
  {NoStop}%
\bibitem [{\citenamefont {Debray}\ \emph {et~al.}(2009)\citenamefont {Debray},
  \citenamefont {Rahman}, \citenamefont {Wan}, \citenamefont {Newrock},
  \citenamefont {Cahay}, \citenamefont {Ngo}, \citenamefont {Ulloa},
  \citenamefont {Herbert}, \citenamefont {Muhammad},\ and\ \citenamefont
  {Johnson}}]{Debray09}%
  \BibitemOpen
  \bibfield  {author} {\bibinfo {author} {\bibfnamefont {P.}~\bibnamefont
  {Debray}}, \bibinfo {author} {\bibfnamefont {S.}~\bibnamefont {Rahman}},
  \bibinfo {author} {\bibfnamefont {J.}~\bibnamefont {Wan}}, \bibinfo {author}
  {\bibfnamefont {R.}~\bibnamefont {Newrock}}, \bibinfo {author} {\bibfnamefont
  {M.}~\bibnamefont {Cahay}}, \bibinfo {author} {\bibfnamefont
  {A.}~\bibnamefont {Ngo}}, \bibinfo {author} {\bibfnamefont {S.}~\bibnamefont
  {Ulloa}}, \bibinfo {author} {\bibfnamefont {S.}~\bibnamefont {Herbert}},
  \bibinfo {author} {\bibfnamefont {M.}~\bibnamefont {Muhammad}}, \ and\
  \bibinfo {author} {\bibfnamefont {M.}~\bibnamefont {Johnson}},\ }\href@noop
  {} {\bibfield  {journal} {\bibinfo  {journal} {Nature Nanotech.}\ }\textbf
  {\bibinfo {volume} {4}},\ \bibinfo {pages} {759} (\bibinfo {year}
  {2009})}\BibitemShut {NoStop}%
\bibitem [{\citenamefont {Nitta}\ \emph {et~al.}(1997)\citenamefont {Nitta},
  \citenamefont {Akazaki}, \citenamefont {Takayanagi},\ and\ \citenamefont
  {Enoki}}]{Nitta97}%
  \BibitemOpen
  \bibfield  {author} {\bibinfo {author} {\bibfnamefont {J.}~\bibnamefont
  {Nitta}}, \bibinfo {author} {\bibfnamefont {T.}~\bibnamefont {Akazaki}},
  \bibinfo {author} {\bibfnamefont {H.}~\bibnamefont {Takayanagi}}, \ and\
  \bibinfo {author} {\bibfnamefont {T.}~\bibnamefont {Enoki}},\ }\href
  {\doibase 10.1103/PhysRevLett.78.1335} {\bibfield  {journal} {\bibinfo
  {journal} {Phys. Rev. Lett.}\ }\textbf {\bibinfo {volume} {78}},\ \bibinfo
  {pages} {1335} (\bibinfo {year} {1997})}\BibitemShut {NoStop}%
\bibitem [{\citenamefont {Winkler}\ \emph {et~al.}(2002)\citenamefont
  {Winkler}, \citenamefont {Noh}, \citenamefont {Tutuc},\ and\ \citenamefont
  {Shayegan}}]{Winkler02}%
  \BibitemOpen
  \bibfield  {author} {\bibinfo {author} {\bibfnamefont {R.}~\bibnamefont
  {Winkler}}, \bibinfo {author} {\bibfnamefont {H.}~\bibnamefont {Noh}},
  \bibinfo {author} {\bibfnamefont {E.}~\bibnamefont {Tutuc}}, \ and\ \bibinfo
  {author} {\bibfnamefont {M.}~\bibnamefont {Shayegan}},\ }\href {\doibase
  10.1103/PhysRevB.65.155303} {\bibfield  {journal} {\bibinfo  {journal} {Phys.
  Rev. B}\ }\textbf {\bibinfo {volume} {65}},\ \bibinfo {pages} {155303}
  (\bibinfo {year} {2002})}\BibitemShut {NoStop}%
\bibitem [{\citenamefont {Bi}\ \emph {et~al.}(2013)\citenamefont {Bi},
  \citenamefont {He}, \citenamefont {Hankiewicz}, \citenamefont {Winkler},
  \citenamefont {Vignale},\ and\ \citenamefont {Culcer}}]{Bi13}%
  \BibitemOpen
  \bibfield  {author} {\bibinfo {author} {\bibfnamefont {X.}~\bibnamefont
  {Bi}}, \bibinfo {author} {\bibfnamefont {P.}~\bibnamefont {He}}, \bibinfo
  {author} {\bibfnamefont {E.~M.}\ \bibnamefont {Hankiewicz}}, \bibinfo
  {author} {\bibfnamefont {R.}~\bibnamefont {Winkler}}, \bibinfo {author}
  {\bibfnamefont {G.}~\bibnamefont {Vignale}}, \ and\ \bibinfo {author}
  {\bibfnamefont {D.}~\bibnamefont {Culcer}},\ }\href {\doibase
  10.1103/PhysRevB.88.035316} {\bibfield  {journal} {\bibinfo  {journal} {Phys.
  Rev. B}\ }\textbf {\bibinfo {volume} {88}},\ \bibinfo {pages} {035316}
  (\bibinfo {year} {2013})}\BibitemShut {NoStop}%
\bibitem [{\citenamefont {Moriya}\ \emph {et~al.}(2014)\citenamefont {Moriya},
  \citenamefont {Sawano}, \citenamefont {Hoshi}, \citenamefont {Masubuchi},
  \citenamefont {Shiraki}, \citenamefont {Wild}, \citenamefont {Neumann},
  \citenamefont {Abstreiter}, \citenamefont {Bougeard}, \citenamefont {Koga},\
  and\ \citenamefont {Machida}}]{Moriya14}%
  \BibitemOpen
  \bibfield  {author} {\bibinfo {author} {\bibfnamefont {R.}~\bibnamefont
  {Moriya}}, \bibinfo {author} {\bibfnamefont {K.}~\bibnamefont {Sawano}},
  \bibinfo {author} {\bibfnamefont {Y.}~\bibnamefont {Hoshi}}, \bibinfo
  {author} {\bibfnamefont {S.}~\bibnamefont {Masubuchi}}, \bibinfo {author}
  {\bibfnamefont {Y.}~\bibnamefont {Shiraki}}, \bibinfo {author} {\bibfnamefont
  {A.}~\bibnamefont {Wild}}, \bibinfo {author} {\bibfnamefont {C.}~\bibnamefont
  {Neumann}}, \bibinfo {author} {\bibfnamefont {G.}~\bibnamefont {Abstreiter}},
  \bibinfo {author} {\bibfnamefont {D.}~\bibnamefont {Bougeard}}, \bibinfo
  {author} {\bibfnamefont {T.}~\bibnamefont {Koga}}, \ and\ \bibinfo {author}
  {\bibfnamefont {T.}~\bibnamefont {Machida}},\ }\href {\doibase
  10.1103/PhysRevLett.113.086601} {\bibfield  {journal} {\bibinfo  {journal}
  {Phys. Rev. Lett.}\ }\textbf {\bibinfo {volume} {113}},\ \bibinfo {pages}
  {086601} (\bibinfo {year} {2014})}\BibitemShut {NoStop}%
\bibitem [{\citenamefont {Winkler}\ \emph {et~al.}(2008)\citenamefont
  {Winkler}, \citenamefont {Culcer}, \citenamefont {Papadakis}, \citenamefont
  {Habib},\ and\ \citenamefont {Shayegan}}]{Winkler08}%
  \BibitemOpen
  \bibfield  {author} {\bibinfo {author} {\bibfnamefont {R.}~\bibnamefont
  {Winkler}}, \bibinfo {author} {\bibfnamefont {D.}~\bibnamefont {Culcer}},
  \bibinfo {author} {\bibfnamefont {S.~J.}\ \bibnamefont {Papadakis}}, \bibinfo
  {author} {\bibfnamefont {B.}~\bibnamefont {Habib}}, \ and\ \bibinfo {author}
  {\bibfnamefont {M.}~\bibnamefont {Shayegan}},\ }\href
  {http://stacks.iop.org/0268-1242/23/i=11/a=114017} {\bibfield  {journal}
  {\bibinfo  {journal} {Semicond. Sci. Technol.}\ }\textbf {\bibinfo {volume}
  {23}},\ \bibinfo {pages} {114017} (\bibinfo {year} {2008})}\BibitemShut
  {NoStop}%
\bibitem [{\citenamefont {Bergmann}(1984)}]{BERGMANN84}%
  \BibitemOpen
  \bibfield  {author} {\bibinfo {author} {\bibfnamefont {G.}~\bibnamefont
  {Bergmann}},\ }\href {\doibase https://doi.org/10.1016/0370-1573(84)90103-0}
  {\bibfield  {journal} {\bibinfo  {journal} {Phys. Rep.}\ }\textbf {\bibinfo
  {volume} {107}},\ \bibinfo {pages} {1 } (\bibinfo {year} {1984})}\BibitemShut
  {NoStop}%
\bibitem [{\citenamefont {Hikami}\ \emph {et~al.}(1980)\citenamefont {Hikami},
  \citenamefont {Larkin},\ and\ \citenamefont {Nagaoka}}]{hikami1980}%
  \BibitemOpen
  \bibfield  {author} {\bibinfo {author} {\bibfnamefont {S.}~\bibnamefont
  {Hikami}}, \bibinfo {author} {\bibfnamefont {A.~I.}\ \bibnamefont {Larkin}},
  \ and\ \bibinfo {author} {\bibfnamefont {Y.}~\bibnamefont {Nagaoka}},\
  }\href@noop {} {\bibfield  {journal} {\bibinfo  {journal} {Progr. Theor. Exp.
  Phys.}\ }\textbf {\bibinfo {volume} {63}},\ \bibinfo {pages} {707} (\bibinfo
  {year} {1980})}\BibitemShut {NoStop}%
\bibitem [{\citenamefont {Iordanskii}\ \emph {et~al.}(1994)\citenamefont
  {Iordanskii}, \citenamefont {Lyanda-Geller},\ and\ \citenamefont
  {Pikus}}]{iordanskii1994}%
  \BibitemOpen
  \bibfield  {author} {\bibinfo {author} {\bibfnamefont {S.}~\bibnamefont
  {Iordanskii}}, \bibinfo {author} {\bibfnamefont {Y.~B.}\ \bibnamefont
  {Lyanda-Geller}}, \ and\ \bibinfo {author} {\bibfnamefont {G.}~\bibnamefont
  {Pikus}},\ }\href@noop {} {\bibfield  {journal} {\bibinfo  {journal} {ZhETF
  Pisma Redaktsiiu}\ }\textbf {\bibinfo {volume} {60}},\ \bibinfo {pages} {199}
  (\bibinfo {year} {1994})}\BibitemShut {NoStop}%
\bibitem [{\citenamefont {Dresselhaus}(1955)}]{Dresselhaus55}%
  \BibitemOpen
  \bibfield  {author} {\bibinfo {author} {\bibfnamefont {G.}~\bibnamefont
  {Dresselhaus}},\ }\href {\doibase 10.1103/PhysRev.100.580} {\bibfield
  {journal} {\bibinfo  {journal} {Phys. Rev.}\ }\textbf {\bibinfo {volume}
  {100}},\ \bibinfo {pages} {580} (\bibinfo {year} {1955})}\BibitemShut
  {NoStop}%
\bibitem [{\citenamefont {Dresselhaus}\ \emph {et~al.}(1992)\citenamefont
  {Dresselhaus}, \citenamefont {Papavassiliou}, \citenamefont {Wheeler},\ and\
  \citenamefont {Sacks}}]{Dresselhaus92}%
  \BibitemOpen
  \bibfield  {author} {\bibinfo {author} {\bibfnamefont {P.~D.}\ \bibnamefont
  {Dresselhaus}}, \bibinfo {author} {\bibfnamefont {C.~M.~A.}\ \bibnamefont
  {Papavassiliou}}, \bibinfo {author} {\bibfnamefont {R.~G.}\ \bibnamefont
  {Wheeler}}, \ and\ \bibinfo {author} {\bibfnamefont {R.~N.}\ \bibnamefont
  {Sacks}},\ }\href {\doibase 10.1103/PhysRevLett.68.106} {\bibfield  {journal}
  {\bibinfo  {journal} {Phys. Rev. Lett.}\ }\textbf {\bibinfo {volume} {68}},\
  \bibinfo {pages} {106} (\bibinfo {year} {1992})}\BibitemShut {NoStop}%
\bibitem [{\citenamefont {Heida}\ \emph {et~al.}(1998)\citenamefont {Heida},
  \citenamefont {van Wees}, \citenamefont {Kuipers}, \citenamefont {Klapwijk},\
  and\ \citenamefont {Borghs}}]{Heida98}%
  \BibitemOpen
  \bibfield  {author} {\bibinfo {author} {\bibfnamefont {J.~P.}\ \bibnamefont
  {Heida}}, \bibinfo {author} {\bibfnamefont {B.~J.}\ \bibnamefont {van Wees}},
  \bibinfo {author} {\bibfnamefont {J.~J.}\ \bibnamefont {Kuipers}}, \bibinfo
  {author} {\bibfnamefont {T.~M.}\ \bibnamefont {Klapwijk}}, \ and\ \bibinfo
  {author} {\bibfnamefont {G.}~\bibnamefont {Borghs}},\ }\href {\doibase
  10.1103/PhysRevB.57.11911} {\bibfield  {journal} {\bibinfo  {journal} {Phys.
  Rev. B}\ }\textbf {\bibinfo {volume} {57}},\ \bibinfo {pages} {11911}
  (\bibinfo {year} {1998})}\BibitemShut {NoStop}%
\bibitem [{\citenamefont {Grundler}(2000)}]{Grundler00}%
  \BibitemOpen
  \bibfield  {author} {\bibinfo {author} {\bibfnamefont {D.}~\bibnamefont
  {Grundler}},\ }\href {\doibase 10.1103/PhysRevLett.84.6074} {\bibfield
  {journal} {\bibinfo  {journal} {Phys. Rev. Lett.}\ }\textbf {\bibinfo
  {volume} {84}},\ \bibinfo {pages} {6074} (\bibinfo {year}
  {2000})}\BibitemShut {NoStop}%
\bibitem [{\citenamefont {Koga}\ \emph
  {et~al.}(2002{\natexlab{b}})\citenamefont {Koga}, \citenamefont {Nitta},
  \citenamefont {Akazaki},\ and\ \citenamefont {Takayanagi}}]{Koga02b}%
  \BibitemOpen
  \bibfield  {author} {\bibinfo {author} {\bibfnamefont {T.}~\bibnamefont
  {Koga}}, \bibinfo {author} {\bibfnamefont {J.}~\bibnamefont {Nitta}},
  \bibinfo {author} {\bibfnamefont {T.}~\bibnamefont {Akazaki}}, \ and\
  \bibinfo {author} {\bibfnamefont {H.}~\bibnamefont {Takayanagi}},\ }\href
  {\doibase 10.1103/PhysRevLett.89.046801} {\bibfield  {journal} {\bibinfo
  {journal} {Phys. Rev. Lett.}\ }\textbf {\bibinfo {volume} {89}},\ \bibinfo
  {pages} {046801} (\bibinfo {year} {2002}{\natexlab{b}})}\BibitemShut
  {NoStop}%
\bibitem [{\citenamefont {Miller}\ \emph {et~al.}(2003)\citenamefont {Miller},
  \citenamefont {Zumb\"uhl}, \citenamefont {Marcus}, \citenamefont
  {Lyanda-Geller}, \citenamefont {Goldhaber-Gordon}, \citenamefont {Campman},\
  and\ \citenamefont {Gossard}}]{Miller03}%
  \BibitemOpen
  \bibfield  {author} {\bibinfo {author} {\bibfnamefont {J.~B.}\ \bibnamefont
  {Miller}}, \bibinfo {author} {\bibfnamefont {D.~M.}\ \bibnamefont
  {Zumb\"uhl}}, \bibinfo {author} {\bibfnamefont {C.~M.}\ \bibnamefont
  {Marcus}}, \bibinfo {author} {\bibfnamefont {Y.~B.}\ \bibnamefont
  {Lyanda-Geller}}, \bibinfo {author} {\bibfnamefont {D.}~\bibnamefont
  {Goldhaber-Gordon}}, \bibinfo {author} {\bibfnamefont {K.}~\bibnamefont
  {Campman}}, \ and\ \bibinfo {author} {\bibfnamefont {A.~C.}\ \bibnamefont
  {Gossard}},\ }\href {\doibase 10.1103/PhysRevLett.90.076807} {\bibfield
  {journal} {\bibinfo  {journal} {Phys. Rev. Lett.}\ }\textbf {\bibinfo
  {volume} {90}},\ \bibinfo {pages} {076807} (\bibinfo {year}
  {2003})}\BibitemShut {NoStop}%
\bibitem [{\citenamefont {Wunderlich}\ \emph {et~al.}(2005)\citenamefont
  {Wunderlich}, \citenamefont {Kaestner}, \citenamefont {Sinova},\ and\
  \citenamefont {Jungwirth}}]{Wunderlich05}%
  \BibitemOpen
  \bibfield  {author} {\bibinfo {author} {\bibfnamefont {J.}~\bibnamefont
  {Wunderlich}}, \bibinfo {author} {\bibfnamefont {B.}~\bibnamefont
  {Kaestner}}, \bibinfo {author} {\bibfnamefont {J.}~\bibnamefont {Sinova}}, \
  and\ \bibinfo {author} {\bibfnamefont {T.}~\bibnamefont {Jungwirth}},\ }\href
  {\doibase 10.1103/PhysRevLett.94.047204} {\bibfield  {journal} {\bibinfo
  {journal} {Phys. Rev. Lett.}\ }\textbf {\bibinfo {volume} {94}},\ \bibinfo
  {pages} {047204} (\bibinfo {year} {2005})}\BibitemShut {NoStop}%
\bibitem [{\citenamefont {Studer}\ \emph {et~al.}(2009)\citenamefont {Studer},
  \citenamefont {Salis}, \citenamefont {Ensslin}, \citenamefont {Driscoll},\
  and\ \citenamefont {Gossard}}]{Studer09}%
  \BibitemOpen
  \bibfield  {author} {\bibinfo {author} {\bibfnamefont {M.}~\bibnamefont
  {Studer}}, \bibinfo {author} {\bibfnamefont {G.}~\bibnamefont {Salis}},
  \bibinfo {author} {\bibfnamefont {K.}~\bibnamefont {Ensslin}}, \bibinfo
  {author} {\bibfnamefont {D.~C.}\ \bibnamefont {Driscoll}}, \ and\ \bibinfo
  {author} {\bibfnamefont {A.~C.}\ \bibnamefont {Gossard}},\ }\href {\doibase
  10.1103/PhysRevLett.103.027201} {\bibfield  {journal} {\bibinfo  {journal}
  {Phys. Rev. Lett.}\ }\textbf {\bibinfo {volume} {103}},\ \bibinfo {pages}
  {027201} (\bibinfo {year} {2009})}\BibitemShut {NoStop}%
\bibitem [{\citenamefont {van Weperen}\ \emph {et~al.}(2015)\citenamefont {van
  Weperen}, \citenamefont {Tarasinski}, \citenamefont {Eeltink}, \citenamefont
  {Pribiag}, \citenamefont {Plissard}, \citenamefont {Bakkers}, \citenamefont
  {Kouwenhoven},\ and\ \citenamefont {Wimmer}}]{vanWeperen15}%
  \BibitemOpen
  \bibfield  {author} {\bibinfo {author} {\bibfnamefont {I.}~\bibnamefont {van
  Weperen}}, \bibinfo {author} {\bibfnamefont {B.}~\bibnamefont {Tarasinski}},
  \bibinfo {author} {\bibfnamefont {D.}~\bibnamefont {Eeltink}}, \bibinfo
  {author} {\bibfnamefont {V.~S.}\ \bibnamefont {Pribiag}}, \bibinfo {author}
  {\bibfnamefont {S.~R.}\ \bibnamefont {Plissard}}, \bibinfo {author}
  {\bibfnamefont {E.~P. A.~M.}\ \bibnamefont {Bakkers}}, \bibinfo {author}
  {\bibfnamefont {L.~P.}\ \bibnamefont {Kouwenhoven}}, \ and\ \bibinfo {author}
  {\bibfnamefont {M.}~\bibnamefont {Wimmer}},\ }\href {\doibase
  10.1103/PhysRevB.91.201413} {\bibfield  {journal} {\bibinfo  {journal} {Phys.
  Rev. B}\ }\textbf {\bibinfo {volume} {91}},\ \bibinfo {pages} {201413}
  (\bibinfo {year} {2015})}\BibitemShut {NoStop}%
\bibitem [{\citenamefont {Morrison}\ \emph {et~al.}(2014)\citenamefont
  {Morrison}, \citenamefont {Wiśniewski}, \citenamefont {Rhead}, \citenamefont
  {Foronda}, \citenamefont {Leadley},\ and\ \citenamefont
  {Myronov}}]{Morrison14}%
  \BibitemOpen
  \bibfield  {author} {\bibinfo {author} {\bibfnamefont {C.}~\bibnamefont
  {Morrison}}, \bibinfo {author} {\bibfnamefont {P.}~\bibnamefont
  {Wiśniewski}}, \bibinfo {author} {\bibfnamefont {S.~D.}\ \bibnamefont
  {Rhead}}, \bibinfo {author} {\bibfnamefont {J.}~\bibnamefont {Foronda}},
  \bibinfo {author} {\bibfnamefont {D.~R.}\ \bibnamefont {Leadley}}, \ and\
  \bibinfo {author} {\bibfnamefont {M.}~\bibnamefont {Myronov}},\ }\href
  {\doibase 10.1063/1.4901107} {\bibfield  {journal} {\bibinfo  {journal}
  {Appl. Phys. Lett.}\ }\textbf {\bibinfo {volume} {105}},\ \bibinfo {pages}
  {182401} (\bibinfo {year} {2014})}\BibitemShut {NoStop}%
\bibitem [{\citenamefont {Failla}\ \emph {et~al.}(2015)\citenamefont {Failla},
  \citenamefont {Myronov}, \citenamefont {Morrison}, \citenamefont {Leadley},\
  and\ \citenamefont {Lloyd-Hughes}}]{Failla15}%
  \BibitemOpen
  \bibfield  {author} {\bibinfo {author} {\bibfnamefont {M.}~\bibnamefont
  {Failla}}, \bibinfo {author} {\bibfnamefont {M.}~\bibnamefont {Myronov}},
  \bibinfo {author} {\bibfnamefont {C.}~\bibnamefont {Morrison}}, \bibinfo
  {author} {\bibfnamefont {D.~R.}\ \bibnamefont {Leadley}}, \ and\ \bibinfo
  {author} {\bibfnamefont {J.}~\bibnamefont {Lloyd-Hughes}},\ }\href {\doibase
  10.1103/PhysRevB.92.045303} {\bibfield  {journal} {\bibinfo  {journal} {Phys.
  Rev. B}\ }\textbf {\bibinfo {volume} {92}},\ \bibinfo {pages} {045303}
  (\bibinfo {year} {2015})}\BibitemShut {NoStop}%
\bibitem [{\citenamefont {Foronda}\ \emph {et~al.}(2015)\citenamefont
  {Foronda}, \citenamefont {Morrison}, \citenamefont {Halpin}, \citenamefont
  {Rhead},\ and\ \citenamefont {Myronov}}]{Foronda15}%
  \BibitemOpen
  \bibfield  {author} {\bibinfo {author} {\bibfnamefont {J.}~\bibnamefont
  {Foronda}}, \bibinfo {author} {\bibfnamefont {C.}~\bibnamefont {Morrison}},
  \bibinfo {author} {\bibfnamefont {J.~E.}\ \bibnamefont {Halpin}}, \bibinfo
  {author} {\bibfnamefont {S.~D.}\ \bibnamefont {Rhead}}, \ and\ \bibinfo
  {author} {\bibfnamefont {M.}~\bibnamefont {Myronov}},\ }\href
  {http://stacks.iop.org/0953-8984/27/i=2/a=022201} {\bibfield  {journal}
  {\bibinfo  {journal} {J. Phys. Condens. Matter}\ }\textbf {\bibinfo {volume}
  {27}},\ \bibinfo {pages} {022201} (\bibinfo {year} {2015})}\BibitemShut
  {NoStop}%
\bibitem [{\citenamefont {Mizokuchi}\ \emph {et~al.}(2017)\citenamefont
  {Mizokuchi}, \citenamefont {Torresani}, \citenamefont {Maurand},
  \citenamefont {Zeng}, \citenamefont {Niquet}, \citenamefont {Myronov},\ and\
  \citenamefont {De~Franceschi}}]{Mizokuchi17}%
  \BibitemOpen
  \bibfield  {author} {\bibinfo {author} {\bibfnamefont {R.}~\bibnamefont
  {Mizokuchi}}, \bibinfo {author} {\bibfnamefont {P.}~\bibnamefont
  {Torresani}}, \bibinfo {author} {\bibfnamefont {R.}~\bibnamefont {Maurand}},
  \bibinfo {author} {\bibfnamefont {Z.}~\bibnamefont {Zeng}}, \bibinfo {author}
  {\bibfnamefont {Y.-M.}\ \bibnamefont {Niquet}}, \bibinfo {author}
  {\bibfnamefont {M.}~\bibnamefont {Myronov}}, \ and\ \bibinfo {author}
  {\bibfnamefont {S.}~\bibnamefont {De~Franceschi}},\ }\href {\doibase
  10.1063/1.4997411} {\bibfield  {journal} {\bibinfo  {journal} {Appl. Phys.
  Lett.}\ }\textbf {\bibinfo {volume} {111}},\ \bibinfo {pages} {063102}
  (\bibinfo {year} {2017})}\BibitemShut {NoStop}%
\bibitem [{\citenamefont {Chuang}\ \emph {et~al.}(2015)\citenamefont {Chuang},
  \citenamefont {Ho}, \citenamefont {Smith}, \citenamefont {Sfigakis},
  \citenamefont {Pepper}, \citenamefont {Chen}, \citenamefont {Fan},
  \citenamefont {Griffiths}, \citenamefont {Farrer}, \citenamefont {Beere},
  \citenamefont {Jones}, \citenamefont {Ritchie},\ and\ \citenamefont
  {Chen}}]{Chuang2015}%
  \BibitemOpen
  \bibfield  {author} {\bibinfo {author} {\bibfnamefont {P.}~\bibnamefont
  {Chuang}}, \bibinfo {author} {\bibfnamefont {S.-C.}\ \bibnamefont {Ho}},
  \bibinfo {author} {\bibfnamefont {L.}~\bibnamefont {Smith}}, \bibinfo
  {author} {\bibfnamefont {F.}~\bibnamefont {Sfigakis}}, \bibinfo {author}
  {\bibfnamefont {M.}~\bibnamefont {Pepper}}, \bibinfo {author} {\bibfnamefont
  {C.-H.}\ \bibnamefont {Chen}}, \bibinfo {author} {\bibfnamefont {J.-C.}\
  \bibnamefont {Fan}}, \bibinfo {author} {\bibfnamefont {J.}~\bibnamefont
  {Griffiths}}, \bibinfo {author} {\bibfnamefont {I.}~\bibnamefont {Farrer}},
  \bibinfo {author} {\bibfnamefont {H.}~\bibnamefont {Beere}}, \bibinfo
  {author} {\bibfnamefont {G.}~\bibnamefont {Jones}}, \bibinfo {author}
  {\bibfnamefont {D.}~\bibnamefont {Ritchie}}, \ and\ \bibinfo {author}
  {\bibfnamefont {T.-M.}\ \bibnamefont {Chen}},\ }\href@noop {} {\bibfield
  {journal} {\bibinfo  {journal} {Nature Nanotech.}\ }\textbf {\bibinfo
  {volume} {10}},\ \bibinfo {pages} {35} (\bibinfo {year} {2015})}\BibitemShut
  {NoStop}%
\bibitem [{\citenamefont {Su}\ \emph {et~al.}(2017)\citenamefont {Su},
  \citenamefont {Chuang}, \citenamefont {Liu}, \citenamefont {Li},\ and\
  \citenamefont {Lu}}]{Su2017}%
  \BibitemOpen
  \bibfield  {author} {\bibinfo {author} {\bibfnamefont {Y.-H.}\ \bibnamefont
  {Su}}, \bibinfo {author} {\bibfnamefont {Y.}~\bibnamefont {Chuang}}, \bibinfo
  {author} {\bibfnamefont {C.-Y.}\ \bibnamefont {Liu}}, \bibinfo {author}
  {\bibfnamefont {J.-Y.}\ \bibnamefont {Li}}, \ and\ \bibinfo {author}
  {\bibfnamefont {T.-M.}\ \bibnamefont {Lu}},\ }\href {\doibase
  10.1103/PhysRevMaterials.1.044601} {\bibfield  {journal} {\bibinfo  {journal}
  {Phys. Rev. Mater.}\ }\textbf {\bibinfo {volume} {1}},\ \bibinfo {pages}
  {044601} (\bibinfo {year} {2017})}\BibitemShut {NoStop}%
\bibitem [{\citenamefont {Glazov}\ and\ \citenamefont
  {Golub}(2006)}]{Glazov2006}%
  \BibitemOpen
  \bibfield  {author} {\bibinfo {author} {\bibfnamefont {M.~M.}\ \bibnamefont
  {Glazov}}\ and\ \bibinfo {author} {\bibfnamefont {L.~E.}\ \bibnamefont
  {Golub}},\ }\href {\doibase 10.1134/S1063782606100150} {\bibfield  {journal}
  {\bibinfo  {journal} {Semiconductors}\ }\textbf {\bibinfo {volume} {40}},\
  \bibinfo {pages} {1209} (\bibinfo {year} {2006})}\BibitemShut {NoStop}%
\bibitem [{\citenamefont {Nakamura}\ \emph {et~al.}(2012)\citenamefont
  {Nakamura}, \citenamefont {Koga},\ and\ \citenamefont {Kimura}}]{Nakamura12}%
  \BibitemOpen
  \bibfield  {author} {\bibinfo {author} {\bibfnamefont {H.}~\bibnamefont
  {Nakamura}}, \bibinfo {author} {\bibfnamefont {T.}~\bibnamefont {Koga}}, \
  and\ \bibinfo {author} {\bibfnamefont {T.}~\bibnamefont {Kimura}},\ }\href
  {\doibase 10.1103/PhysRevLett.108.206601} {\bibfield  {journal} {\bibinfo
  {journal} {Phys. Rev. Lett.}\ }\textbf {\bibinfo {volume} {108}},\ \bibinfo
  {pages} {206601} (\bibinfo {year} {2012})}\BibitemShut {NoStop}%
\bibitem [{\citenamefont {Liang}\ \emph {et~al.}(2015)\citenamefont {Liang},
  \citenamefont {Cheng}, \citenamefont {Wei}, \citenamefont {Luo},
  \citenamefont {Yu}, \citenamefont {Zeng},\ and\ \citenamefont
  {Zhang}}]{Liang15}%
  \BibitemOpen
  \bibfield  {author} {\bibinfo {author} {\bibfnamefont {H.}~\bibnamefont
  {Liang}}, \bibinfo {author} {\bibfnamefont {L.}~\bibnamefont {Cheng}},
  \bibinfo {author} {\bibfnamefont {L.}~\bibnamefont {Wei}}, \bibinfo {author}
  {\bibfnamefont {Z.}~\bibnamefont {Luo}}, \bibinfo {author} {\bibfnamefont
  {G.}~\bibnamefont {Yu}}, \bibinfo {author} {\bibfnamefont {C.}~\bibnamefont
  {Zeng}}, \ and\ \bibinfo {author} {\bibfnamefont {Z.}~\bibnamefont {Zhang}},\
  }\href {\doibase 10.1103/PhysRevB.92.075309} {\bibfield  {journal} {\bibinfo
  {journal} {Phys. Rev. B}\ }\textbf {\bibinfo {volume} {92}},\ \bibinfo
  {pages} {075309} (\bibinfo {year} {2015})}\BibitemShut {NoStop}%
\bibitem [{\citenamefont {Grbi\ifmmode~\acute{c}\else \'{c}\fi{}}\ \emph
  {et~al.}(2008)\citenamefont {Grbi\ifmmode~\acute{c}\else \'{c}\fi{}},
  \citenamefont {Leturcq}, \citenamefont {Ihn}, \citenamefont {Ensslin},
  \citenamefont {Reuter},\ and\ \citenamefont {Wieck}}]{Grbic08}%
  \BibitemOpen
  \bibfield  {author} {\bibinfo {author} {\bibfnamefont {B.}~\bibnamefont
  {Grbi\ifmmode~\acute{c}\else \'{c}\fi{}}}, \bibinfo {author} {\bibfnamefont
  {R.}~\bibnamefont {Leturcq}}, \bibinfo {author} {\bibfnamefont
  {T.}~\bibnamefont {Ihn}}, \bibinfo {author} {\bibfnamefont {K.}~\bibnamefont
  {Ensslin}}, \bibinfo {author} {\bibfnamefont {D.}~\bibnamefont {Reuter}}, \
  and\ \bibinfo {author} {\bibfnamefont {A.~D.}\ \bibnamefont {Wieck}},\ }\href
  {\doibase 10.1103/PhysRevB.77.125312} {\bibfield  {journal} {\bibinfo
  {journal} {Phys. Rev. B}\ }\textbf {\bibinfo {volume} {77}},\ \bibinfo
  {pages} {125312} (\bibinfo {year} {2008})}\BibitemShut {NoStop}%
\bibitem [{\citenamefont {Sugahara}\ and\ \citenamefont
  {Nitta}(2010)}]{sugahara2010}%
  \BibitemOpen
  \bibfield  {author} {\bibinfo {author} {\bibfnamefont {S.}~\bibnamefont
  {Sugahara}}\ and\ \bibinfo {author} {\bibfnamefont {J.}~\bibnamefont
  {Nitta}},\ }\href@noop {} {\bibfield  {journal} {\bibinfo  {journal}
  {‎Proc. IEEE}\ }\textbf {\bibinfo {volume} {98}},\ \bibinfo {pages} {2124}
  (\bibinfo {year} {2010})}\BibitemShut {NoStop}%
\bibitem [{\citenamefont {Koo}\ \emph {et~al.}(2009)\citenamefont {Koo},
  \citenamefont {Kwon}, \citenamefont {Eom}, \citenamefont {Chang},
  \citenamefont {Han},\ and\ \citenamefont {Johnson}}]{koo2009}%
  \BibitemOpen
  \bibfield  {author} {\bibinfo {author} {\bibfnamefont {H.~C.}\ \bibnamefont
  {Koo}}, \bibinfo {author} {\bibfnamefont {J.~H.}\ \bibnamefont {Kwon}},
  \bibinfo {author} {\bibfnamefont {J.}~\bibnamefont {Eom}}, \bibinfo {author}
  {\bibfnamefont {J.}~\bibnamefont {Chang}}, \bibinfo {author} {\bibfnamefont
  {S.~H.}\ \bibnamefont {Han}}, \ and\ \bibinfo {author} {\bibfnamefont
  {M.}~\bibnamefont {Johnson}},\ }\href@noop {} {\bibfield  {journal} {\bibinfo
   {journal} {Science}\ }\textbf {\bibinfo {volume} {325}},\ \bibinfo {pages}
  {1515} (\bibinfo {year} {2009})}\BibitemShut {NoStop}%
\bibitem [{\citenamefont {Winkler}\ \emph {et~al.}(2003)\citenamefont
  {Winkler}, \citenamefont {Papadakis}, \citenamefont {De~Poortere},\ and\
  \citenamefont {Shayegan}}]{Winkler2003book}%
  \BibitemOpen
  \bibfield  {author} {\bibinfo {author} {\bibfnamefont {R.}~\bibnamefont
  {Winkler}}, \bibinfo {author} {\bibfnamefont {S.}~\bibnamefont {Papadakis}},
  \bibinfo {author} {\bibfnamefont {E.}~\bibnamefont {De~Poortere}}, \ and\
  \bibinfo {author} {\bibfnamefont {M.}~\bibnamefont {Shayegan}},\ }\href@noop
  {} {\emph {\bibinfo {title} {Spin-Orbit Coupling in Two-Dimensional Electron
  and Hole Systems}}},\ Vol.~\bibinfo {volume} {41}\ (\bibinfo  {publisher}
  {Springer},\ \bibinfo {year} {2003})\BibitemShut {NoStop}%
\bibitem [{\citenamefont {Habib}\ \emph {et~al.}(2004)\citenamefont {Habib},
  \citenamefont {Tutuc}, \citenamefont {Melinte}, \citenamefont {Shayegan},
  \citenamefont {Wasserman}, \citenamefont {Lyon},\ and\ \citenamefont
  {Winkler}}]{Habib04}%
  \BibitemOpen
  \bibfield  {author} {\bibinfo {author} {\bibfnamefont {B.}~\bibnamefont
  {Habib}}, \bibinfo {author} {\bibfnamefont {E.}~\bibnamefont {Tutuc}},
  \bibinfo {author} {\bibfnamefont {S.}~\bibnamefont {Melinte}}, \bibinfo
  {author} {\bibfnamefont {M.}~\bibnamefont {Shayegan}}, \bibinfo {author}
  {\bibfnamefont {D.}~\bibnamefont {Wasserman}}, \bibinfo {author}
  {\bibfnamefont {S.~A.}\ \bibnamefont {Lyon}}, \ and\ \bibinfo {author}
  {\bibfnamefont {R.}~\bibnamefont {Winkler}},\ }\href {\doibase
  10.1063/1.1806543} {\bibfield  {journal} {\bibinfo  {journal} {Appl. Phys.
  Lett.}\ }\textbf {\bibinfo {volume} {85}},\ \bibinfo {pages} {3151} (\bibinfo
  {year} {2004})}\BibitemShut {NoStop}%
\end{thebibliography}


\end{document}